\documentclass[journal=nalefd,
manuscript=article,
 amsmath,amssymb,
]{achemso}
\usepackage{mathrsfs}
\usepackage{graphicx}
\usepackage{dcolumn}
\usepackage{bm}
\usepackage{amsmath,mathrsfs}    
\usepackage{hyperref}   
\usepackage{upgreek}
\usepackage{textcomp}
\usepackage[T1]{fontenc}

\title{Real-time detection of every Auger recombination in a self-assembled quantum dot}
\author{P.~Lochner}
\author{A.~Kurzmann}
 \altaffiliation{Solid State Physics Laboratory, ETH Zurich, Otto-Stern-Weg~1, 8093~Zurich, Switzerland}
\author{J.~Kerski}
\author{P.~Stegmann}
\author{J.~K\"onig}
 \affiliation{University of Duisburg-Essen, Faculty of Physics and CENIDE, Lotharstr.~1, 47057~Duisburg, Germany}
\author{A.~D.~Wieck}
\author{A.~Ludwig}
 \affiliation{Lehrstuhl f\"ur Angewandte Festk\"orperphysik, Ruhr-Universit\"at Bochum, Universit\"atsstra{\ss}e~150, 44780~Bochum, Germany}
\author{A.~Lorke}
\author{M.~Geller}
 \email{martin.geller@uni-due.de}
 \affiliation{University of Duisburg-Essen, Faculty of Physics and CENIDE, Lotharstr.~1, 47057~Duisburg, Germany}

\date{\today}

\begin{document}
\begin{abstract}
Auger recombination is a non-radiative process, where the recombination energy of an electron-hole pair is transferred to a third charge carrier.  It is a common effect in colloidal quantum dots that quenches the radiative emission with an Auger recombination time below nanoseconds. In self-assembled QDs, the Auger recombination has been observed with a much longer recombination time in the order of microseconds. Here, we use two-color laser excitation on the exciton and trion transition in resonance fluorescence on a single self-assembled quantum dot to monitor in real-time every quantum event of the Auger process. Full counting statistics on the random telegraph signal give access to the cumulants and  demonstrate the tunability of the Fano factor from a Poissonian to a sub-Poissonian distribution by Auger-mediated electron emission from the dot. Therefore, the Auger process can be used to tune optically the charge carrier occupation of the dot by the incident laser intensity; independently from the electron tunneling from the reservoir by the gate voltage. Our findings are not only highly relevant for the understanding of the Auger process, it also demonstrates the perspective of the Auger effect for controlling precisely the charge state in a quantum system by optical means.  

\end{abstract}

\maketitle


The excitonic transitions in self-assembled quantum dots (QDs) \cite{Bimberg1999,Petroff2001} realize perfectly a two-level system in a solid-state environment. These transitions can be used to generate single photon sources \cite{Michler2000,Yuan2001} with high photon indistinguishability\cite{Santori2002,Matthiesen2013}, an important prerequisite to use quantum dots as building blocks in (optical) quantum information and communication technologies\cite{Kimble2008,Ladd2010}. Moreover, self-assembled QDs are still one of the best model systems to study in an artificial atom the carrier dynamics\cite{Kurzmann2016b,Geller2019}, the spin- and angular-momentum properties\cite{Bayer2000,Vamivakas2009} and charge carrier interactions\cite{Labud2014}. One important effect of carrier interactions is the Auger process: An electron-hole pair recombines and instead of emitting a photon, the recombination energy is transferred to a third charge carrier, which is then energetically ejected from the QD\cite{Kharchenko1996,Efros1997,Fisher2005,Jha2009}. This is a common effect, mostly studied in colloidal QDs, where it quenches the radiative emission with recombination times in the order of picoseconds to nanoseconds\cite{Vaxenburg2015,Klimov2000,Park2014}. This limits the efficiency of optical devices containing QDs like LEDs\cite{Caruge2008,Cho2009} or single photon sources\cite{Brokmann2004,Michler2000a,Lounis2000}. In self-assembled QDs, Auger recombination was speculated to be absent, and only recently, it was directly observed in optical measurements on a single self-assembled QD coupled to a charge reservoir with recombination times in the order of microseconds\cite{Kurzmann2016}. As a single Auger process is a quantum event, it is unpredictable and only the statistical evaluation of many processes gives access to the physical information of the recombination process\cite{Levitov1996,Blanter2000}. The most in-depth evaluation - the so-called full counting statistics - becomes possible when each single quantum event in a time trace is recorded. Such real-time detection in optical experiments on a single self-assembled QD have until now only been shown for the statistical process of electron tunneling between the QD and a charge reservoir, where tunneling and spin-flip rates could be tuned by the applied electric and magnetic field\cite{Kurzmann2019}. 

Here, Auger recombination in a single self-assembled QD is investigated by optical real-time measurements of the random telegraph signal. With the technique of two-laser excitation, we are able to detect every single quantum event of the Auger recombination. These events take place in the single QD, leaving the quantum dot empty until single-electron tunneling into the QD from the charge reservoir takes place again. This reservoir is coupled to the QD with a small tunneling rate in the order of ms$^{-1}$. The laser intensity, exciting the trion transition, precisely controls the electron emission by the Auger recombination and, hence, the average occupation with an electron. It also tunes the Fano factor from a Poissonian to a sub-Poissonian distribution, which we observe in analyzing the random telegraph signal by methods of full counting statistics. 

The investigated sample was grown by molecular beam epitaxy (MBE) with a single layer of self-assembled In(Ga)As QDs embedded in a p-i-n diode (see Supporting Information for details). A highly n-doped GaAs layer acts as charge reservoir, which is coupled to the QDs via a tunneling barrier, while a highly p-doped GaAs layer defines an epitaxial gate\cite{Ludwig2017}.  An applied gate voltage $V_\text{G}$ shifts energetically the QD states with respect to the Fermi energy in the electron reservoir and controls the charge state of the dots by electron tunneling through the tunneling barrier. The sample is integrated into a confocal microscope setup within a bath cryostat at 4.2\,K for resonant fluorescence (RF) measurements (see Methods). 

\begin{figure*}
	\centering
	\includegraphics[width=1\columnwidth]{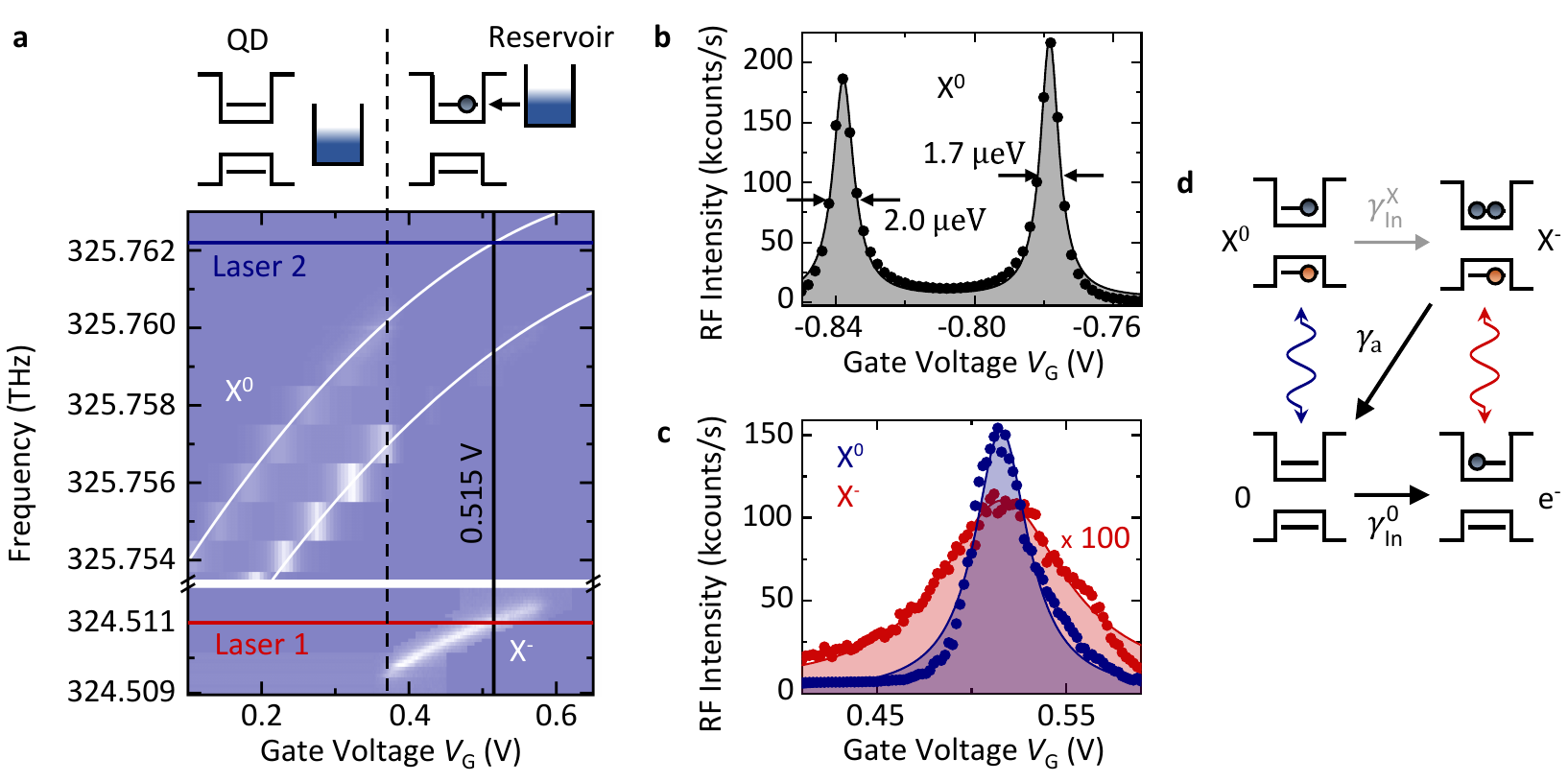}
	\caption{\textbf{Resonance fluorescence (RF) of the exciton (X\textsuperscript{0}) and trion (X\textsuperscript{-}) transition.} \textbf{a} At a gate voltage of $V_\text{G}=0.375\,\text{V}$ (vertical dashed line), the electron ground state is in resonance with the Fermi energy in the charge reservoir. The exciton transition vanishes and the trion transition can be excited at lower frequencies. \textbf{b} Gate voltage scan of the exciton at a fixed excitation frequency of 325.710\,THz (laser excitation intensity of $1.6\cdot10^{-3}\,\upmu$W/$\upmu$m$^2$). \textbf{c} Two-color laser excitation  with laser~1 and laser~2 (red and blue line in panel \textbf{a}) shows a bright exciton fluorescence X\textsuperscript{0} at a gate voltage where, in equilibrium, an electron would occupy the QD and quenches the X\textsuperscript{0} transition. X\textsuperscript{-} signal is scaled up by factor 100. Simultaneous excitation of the trion transition X\textsuperscript{-} empties the dot by Auger recombination and the exciton transition can be excited with the second laser until an electron tunnels into the dot again (see panel \textbf{d} for a schematic representation).}
	\label{1}
\end{figure*}

Figure~\ref{1} shows the RF of the neutral exciton (X\textsuperscript{0}) and the negatively charged exciton, called trion (X\textsuperscript{-}). A RF measurement as function of gate voltage in Figure~\ref{1}\textbf{b} shows the fine-structure split exciton\cite{Hoegele2004} with an average linewidth of about 1.8\,$\upmu$eV at low excitation intensity ($1.6\cdot10^{-3}\,\upmu$W/$\upmu$m$^2$). Please note, that this measurement was recorded at a laser energy where the exciton gets into resonance at negative gate voltages because here, the measurement conditions were the best. The quantum-confined Stark effect shifts the exciton resonance X\textsuperscript{0} for higher gate voltages to higher frequencies up to 325.760\,THz, seen in Figure~\ref{1}\textbf{a}. This quadratic Stark shift of the two exciton transitions\cite{Li2000} is indicated by two white lines. At a voltage of about 0.375\,V (dashed vertical line in Fig.~\ref{1}\textbf{a}), the electron ground state in the dot is in resonance with the Fermi energy in the charge reservoir. An electron tunnels into the QD and the exciton transition vanishes while the trion transition can be excited at lower frequencies from 324.5095\,THz to 324.5115\,THz. 

The spectrum of the exciton (blue dots) and the trion transition (red dots) under two-laser excitation is shown in Figure~\ref{1}\textbf{c}. The trion transition is measured at a laser frequency of 324.511\,THz (corresponding to the red line, "Laser~1" in Fig.~\ref{1}\textbf{a}) and a laser excitation intensity of $8\cdot10^{-6}\,\upmu$W/$\upmu$m$^2$ at a gate voltage of 0.515\,V. The exciton spectrum in Figure~\ref{1}\textbf{c} was obtained simultaneously by a second laser 2 on the exciton transition (blue line in Fig.~\ref{1}\textbf{a} at 325.7622\,THz) with a laser excitation intensity of $1.6\cdot10^{-3}\,\upmu$W/$\upmu$m$^2$, as the Auger recombination with rate $\gamma_\text{a}$ leads to an empty QD until an electron tunnels into the dot from the reservoir with rate $\gamma_\text{In}=\gamma_\text{In}^0+\gamma_\text{In}^\text{X}$. This rate comprises the tunneling into the empty dot $\gamma_\text{In}^0$ and the tunneling into the dot charged with an exciton $\gamma_\text{In}^\text{X}$\cite{Seidl2005} (see Fig.~\ref{1}\textbf{d} for a schematic representation). This has been explained previously in Kurzmann et al.\cite{Kurzmann2016} with the important conclusion that the intensity ratio between trion/exciton intensity in equilibrium measurements is given by the ratio between Auger/tunneling rate $\gamma_\text{a}/\gamma_\text{In}$. As the tunneling rate $\gamma_\text{In}$ in the sample used here is in the range of ms$^{-1}$, the Auger rate $\gamma_\text{a}$ exceeds the tunneling rate by more than two orders of magnitude (see below). As a consequence, the intensity of the trion transition in equilibrium is by more than two orders of magnitude smaller than the exciton transition.

\begin{figure*}
	\centering
	\includegraphics[width=0.5\columnwidth]{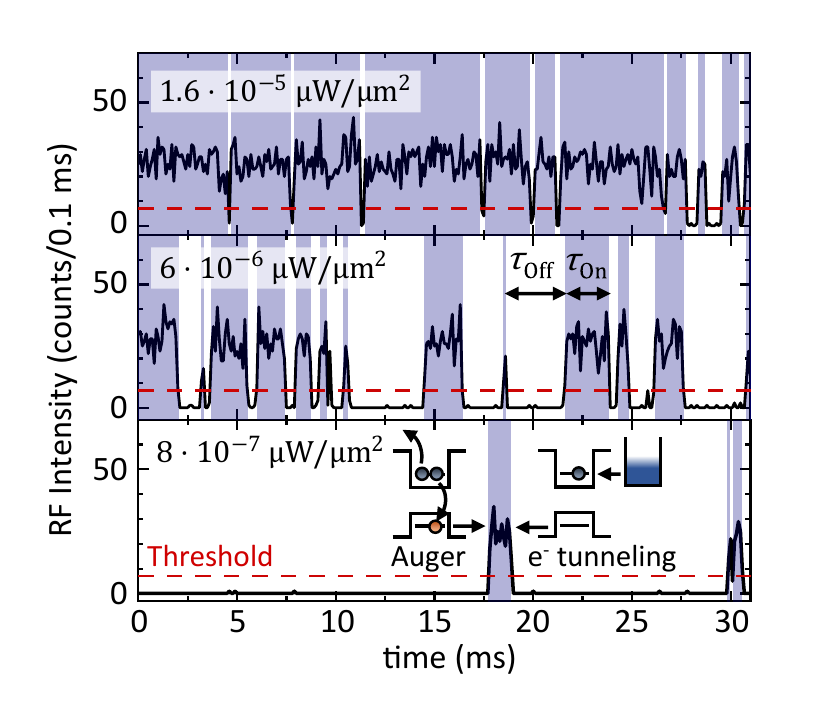}
	\caption{\textbf{Time-resolved RF random telegraph signal.} At a gate voltage of 0.515\,V, the trion and exciton are excited simultaneously (like in Fig.~\ref{1}\textbf{c}). The intensity of the exciton excitation laser 2 is held constant at $1.6\cdot10^{-3}\,\upmu$W/$\upmu$m$^2$, which is far below the saturation of the RF signal (see Supporting Information). The intensity of the trion excitation laser 1 is varied (from top to bottom: $1.6\cdot10^{-5}$,  $6\cdot10^{-6}$ and $8\cdot10^{-7}\,\upmu$W/$\upmu$m$^2$). Every time, an Auger recombination takes place, the dot is emptied, and exciton RF signal turns on. After a time $\tau_\text{On}$, an electron tunnels into the QD and the exciton RF signal quenches. All intensities smaller than the threshold (red line at 7\,counts/0.1\,ms) are counted as "exciton off" (white areas), all intensities above the threshold are counted as "exciton on" (blue areas).} 
	\label{2}
\end{figure*}

The interplay between electron tunneling and optical-driven Auger recombination can be studied in more detail by a real-time random telegraph signal of the resonance fluorescence. In these measurements, the time stamp of every detected RF photon is recorded, see Figure~\ref{2}, enabling the evaluation by full counting statistics. As the intensity of the trion is very weak, the random telegraph signal has been investigated in a two-color excitation scheme. The bright exciton transition with count rates exceeding 10\,MCounts/s (see Supporting Information) is used as an optical detector for the telegraph signal of the Auger recombination. In this two-color laser excitation scheme, the "exciton off" signal corresponds to the "trion on" signal and vice versa\cite{Kurzmann2016}. Hence, the trion statistics can directly be determined from the "inverse" exciton signal. The intensity of the exciton excitation laser~2 is held constant at $1.6\cdot 10^{-3}\,\upmu$W/$\upmu$m$^2$. This intensity is far below the saturation of the RF signal of the exciton (see Supporting Information) and avoids the photon-induced electron capture at high excitation intensities\cite{Kurzmann2016a}. However, this laser intensity yields count rates above 200\,kcounts/s (see Fig.~\ref{1}\textbf{b}), sufficiently-high for recording single quantum events in a real-time measurement\cite{Kurzmann2019}. While the intensity of the exciton detection laser 2 is kept constant, the laser intensity of the trion excitation laser~1 is increased from $1.6\cdot10^{-7}\,\upmu$W/$\upmu$m$^2$ up to $1.6\cdot10^{-5}\,\upmu$W/$\upmu$m$^2$.

For every trion laser intensity, the time-resolved RF signal is recorded for 15 minutes using a fast (350\,ps) avalanche photo diode and a bin time of 100\,$\upmu$s. Figure \ref{2} shows parts of three different time traces at three different trion laser 1 intensities. As the exciton laser 2 intensity always exceeds the trion laser 1 intensity by at least nearly two orders of magnitude, the small amount of RF counts from the trion can be neglected. As a consequence, the detected RF signal of the exciton is directly related to the Auger recombination: An Auger recombination empties the dot and the exciton transition detects an empty dot  (no trion transition possible) with a count rate of about 25~counts per bin time (100\,$\upmu$s). After a time $\tau_\text{On}$, an electron tunnels into the QD in Figure~\ref{2} and the exciton RF signal quenches for a charged dot (trion transition possible) until, after a time $\tau_\text{Off}$, another Auger recombination happens. 

Increasing the trion laser intensity from $8\cdot10^{-7}$ up to $1.6\cdot10^{-5}\,\upmu$W/$\upmu$m$^2$ in Figure~\ref{2} increases the probability of an electron emission with rate $\gamma_\text{E}= n \gamma_\text{a}$ by an Auger process, as the probability for occupation of the dot with a trion $n$ increases with increasing laser 1 intensity. Therefore, the exciton transition is observed most frequently for the highest trion laser intensity. This can be observed in Figure~\ref{2}, where the optical random telegraph signal is compared for three different trion excitation intensities. A threshold between exciton "on" and "off" is set for the following statistical evaluation\cite{Gustavsson2009,Gustavsson2006}. All exciton RF intensities smaller than this threshold (dashed red line at 7\,counts/0.1\,ms in Fig.~\ref{2}) are counted as "exciton off" (white areas), all intensities above the threshold are counted as "exciton on" (blue areas). 

\begin{figure*}
	\centering
	\includegraphics[width=\columnwidth]{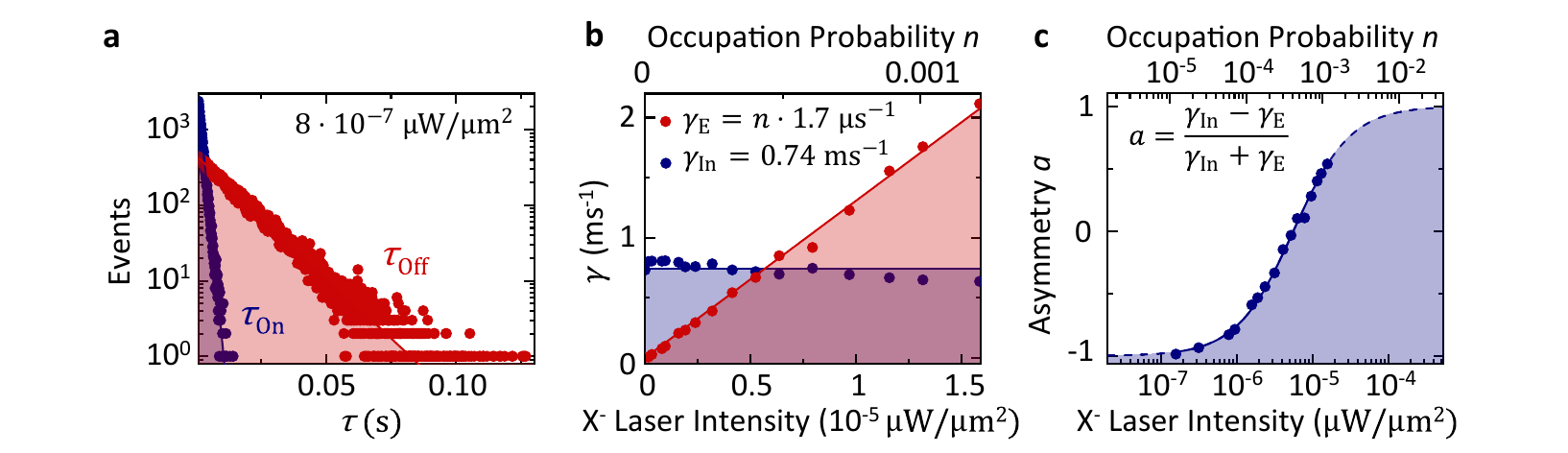}
	\caption{\textbf{Auger and tunneling rates from the time-resolved RF random telegraph signal.} Panel \textbf{a} shows the probability distribution of the "off" (red dots) and "on"-times (blue dots) from the measurement at $8\cdot10^{-7}\,\upmu$W/$\upmu$m$^2$ trion laser excitation intensity (laser~1). Fitting these data yields the emission rate $\gamma_\text{E}=n\gamma_\text{a}$ and the tunneling rate of an electron into the dot $\gamma_\text{In}$. In panel \textbf{b}, the emission and tunneling rates are plotted as function of the trion laser intensity (red and blue dots, respectively). The tunneling rate remains constant at a mean value of 0.74\,ms$^{-1}$, the emission rate increases linearly with the laser intensity and, accordingly, with the occupation probability $n$ (top axis). \textbf{c} Asymmetry $a$, calculated from the emission and tunneling rates as shown in the inset, as a function of the trion laser intensity.}
	\label{3}
\end{figure*}

From these time-resolved RF data sets, the Auger and tunneling rates can be determined by analysing the probability distributions of the "off"-times $\tau_\text{Off}$ and the "on"-times $\tau_\text{On}$ for every 15~minutes long data set\cite{Gustavsson2009}. A representative distribution at a trion laser intensity of $8\cdot10^{-7}\,\upmu$W/$\upmu$m$^2$ can be seen in Figure~\ref{3}\textbf{a}. An exponential fit to the "on"-times (blue line in Fig.~\ref{3}\textbf{a}) yields the tunneling rate $\gamma_\text{In}$ into the QD, while an exponential fit to the "off"-times (red line in Fig.~\ref{3}\textbf{a}) yields the emission rate $\gamma_\text{E}=n\gamma_\text{a}$ for this specific trion laser 1 intensity. In the example in Figure~\ref{3}\textbf{a}, we find $\gamma_\text{In}=0.80\,$ms$^{-1}$ and $\gamma_\text{E}=0.074\,$ms$^{-1}$. As discussed above, the probability for emitting an electron by an Auger recombination process increases with the occupation probability of the QD with a trion $n$. 

The occupation probability with a trion $n$ depends on the laser 1 excitation intensity and has been determined from a pulsed measurement of the trion RF intensity, where the highest trion intensity corresponds to an occupation probability of $n=$ 0.5\cite{Loudon2000} (see Supporting Information for more details). Figure~\ref{3}\textbf{b} shows the expected linear dependence of the electron emission rate $\gamma_\text{E}=n \gamma_\text{a}$ on the occupation probability of the QD with a trion $n$; tuning the emission rate $\gamma_\text{E}$ from almost zero to more than $\gamma_\text{E}= 2$ ms$^{-1}$. The Auger rate is the proportional factor $\gamma_\text{a}$=1.7\,$\upmu\text{s}^{-1}$ (red data points) and in good agreement with the value obtained before for a different QD with slightly different size\cite{Kurzmann2016}. The tunneling rate $\gamma_\text{In}$ remains approximately constant at a mean value of 0.74\,ms$^{-1}$ (blue data points in Fig.~\ref{3}\textbf{b}). This is in agreement with the probability for an electron to tunnel into the empty QD at a constant gate voltage: it is independent on the trion laser intensity. That means, we are able to use the Auger recombination to tune optically the electron emission rate independently from the gate voltage, influencing the emission rate without changing the rate for capturing an electron into the QD (here by the tunneling rate $\gamma_\text{In}$). An independent tuning of electron emission and capture rate is usually not possible for a QD that is tunnel-coupled to one charge reservoir. Changing the coupling strength or Fermi energy by a gate voltage always changes both rates for tunneling into and out of the dot simultaneously.  

Using the standard methods of full counting statistics\cite{Gustavsson2009,Flindt2009} in the following, first of all the asymmetry $a=\frac{\gamma_\text{In}-\gamma_\text{E}}{\gamma_\text{In}+\gamma_\text{E}}$ between the tunneling $\gamma_\text{In}$ and emission rate $\gamma_\text{E}$ has been evaluated. The asymmetry in Figure~\ref{3}\textbf{c} can be tuned by the trion excitation laser intensity from -1 up to 0.55 at a maximum laser intensity of $1.6\cdot10^{-5}\,\upmu$W/$\upmu$m$^2$. Important to mention here: At high trion laser intensities above $1.6\cdot10^{-5}\,\upmu$W/$\upmu$m$^2$, the electron emission by Auger recombination after an tunneling event from the reservoir happens much faster than the bin time of 0.1\,ms. Therefore, the RF intensity within the bin time is not falling below the threshold and these events are not detected, i.\,e.~the maximum bandwidth of 10\,kHz (given by the bin time) of the optical detection scheme distorts the statistical analysis at trion laser intensities above $1.6\cdot10^{-5}\,\upmu$W/$\upmu$m$^2$. Below this laser intensity, every single Auger recombination event is detected in the real-time telegraph signal.

\begin{figure}
	\centering
	\includegraphics[width=\columnwidth]{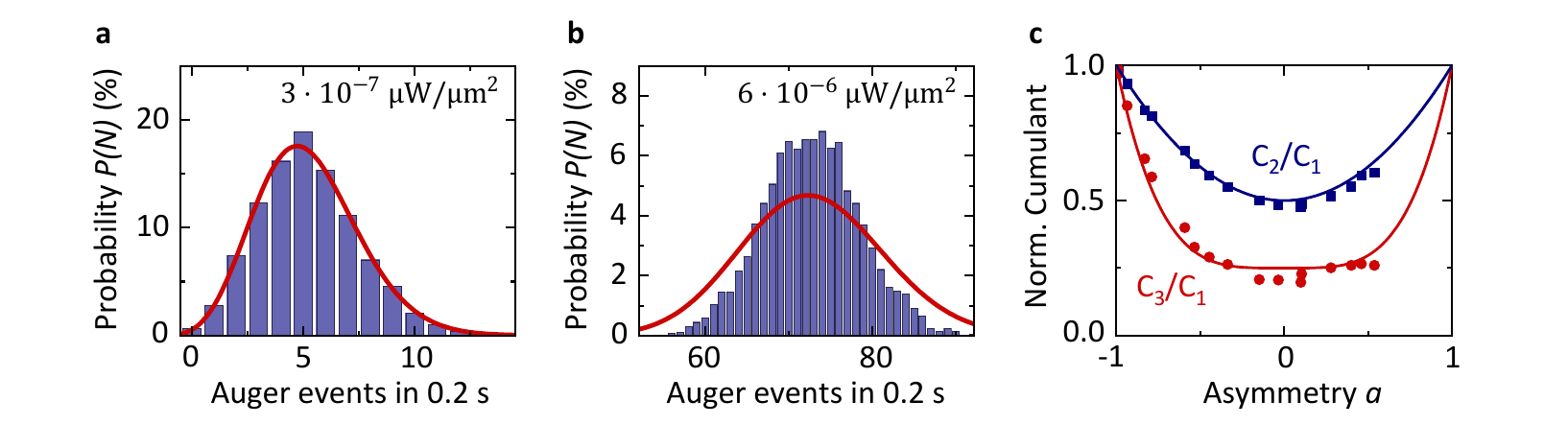}
	\caption{\textbf{Probability distribution and cumulants of the time-resolved RF random telegraph signal.} Panel \textbf{a} and \textbf{b} show the probability $P(N)$ for a number $N$ of Auger events in a bin time of 200\,ms (blue bars) and the Poissonian distribution related to the mean value of the probability $P(N)$ (red curve). At a trion excitation intensity (laser~1) of $3\cdot10^{-7}\,\upmu$W/$\upmu$m$^2$ (panel \textbf{a}), which corresponds to an asymmetry close to -1, the probability $P(N)$ is close to a Poissonian distribution. At a trion excitation intensity of $6\cdot10^{-6}\,\upmu$W/$\upmu$m$^2$ (panel \textbf{b}), which corresponds to an asymmetry close to 0, the probability $P(N)$ is sub-Poissonian. Panel \textbf{c} shows the second (blue) and third (red) normalized cumulant as a function of the asymmetry. Symbols are measured values, lines are calculated curves for a two-state system\cite{Gustavsson2006}.}
	\label{4}
\end{figure}

Finally, full counting statistics\cite{Gustavsson2009,Fricke2007,Gorman2017} is performed on the telegraph signal: Every 15-min long telegraph signal is divided into sections with length $t_0$. The number $N$ of Auger events within the time interval $t_0$ is counted. Figure~\ref{4}\textbf{a} and \textbf{b} show two examples for the corresponding probability distributions $P(N)$ in the limit of large $t_0$ (0.2\,s). At an asymmetry close to -1 (a trion laser intensity of $3\cdot 10^{-7}\,\upmu$W/$\upmu$m$^2$, Fig.~\ref{4}\textbf{a}), the probability is close to a Poissonian distribution. At an asymmetry of about 0 (laser intensity of $6\cdot 10^{-6}\,\upmu$W/$\upmu$m$^2$, Fig.~\ref{4}\textbf{b}), the probability distribution is sub-Poissonian, indicating a relation between Auger recombination and electron tunneling. The Auger recombination emits an electron after an electron has tunneled from the reservoir into the dot. Vice versa, the electron can only tunnel after the Auger recombination has emptied the QD. From the probability distributions, the cumulants $C_m(t_0)=\partial_z^m \ln \mathcal{M}(z,t_0)|_{z=0}$ can be derived with the generating function $\mathcal{M}(z,t_0)=\sum_{N}e^{zN}P(N)$\cite{Gustavsson2009}. The first cumulant $C_1$ corresponds to the mean value, the second cumulant $C_2$ is the variance and the third one describes the skewness of the distribution. The second and third normalized cumulant in the limit of large $t_0$ (20\,ms and 5\,ms, respectively) can be seen as data points in Figure~\ref{4}\textbf{c}. For a two-state system, theory predicts for these normalized cumulants in the long-time limit $C_2/C_1=(1+a^2)/2$ (also called "Fano factor") and $C_3/C_1=(1+3a^4)/4$\cite{Gustavsson2006}, shown as lines in Figure~\ref{4}\textbf{c}. The data for the second and third normalized cumulant coincide perfectly with the calculated curves. We can conclude from the statistical analysis that the QD behaves like a two-state system, where one state is the QD charged with one electron (or a trion after optical excitation) and the other state is the empty dot (or charged with an exciton). The QD charged with one electron cannot be distinguished from the dot containing a trion (same for empty dot and exciton) as the optical transition times in the order of nanoseconds are orders of magnitude faster than the tunneling and emission time by the Auger recombination \cite{Zrenner2002}. The statistical analysis demonstrates the influence of the Auger recombination on the cumulants, especially on the Fano factor, which can be tune from $F=1$ to $F=0.5$ by increasing the incident laser intensity on the trion transition.

In summary, we performed real-time RF random telegraph measurements and studied full counting statistics of the Auger effect in a single self-assembled QD. With this technique, we were able to measure every single Auger recombination as a quantum jump from a charged to an uncharged QD; followed by single-electron tunneling. The full counting statistics gives access to the normalized cumulants and demonstrates the tunability of the Fano factor from Possonian to sub-Poissonian distribution by the incident laser intensity on the trion transition. Comparison with theoretical prediction shows that the empty and charged QD with the Auger recombination and tunneling follows a dynamical two-state system. For future quantum state preparation, the Auger process can be used to control optically the charge state in a quantum system by optical means.

\section{Methods}

As the same measurement technique is used, this methods section follows the Supplemental information of Kurzmann et al.\cite{Kurzmann2019}. 

\subsection{Optical measurements}

Resonant optical excitation and collection of the fluorescence light is used to detect the optical response of the single self-assembled QD, where the resonance condition is achieved by applying a specific gate voltage between the gate electrode and the Ohmic back contact. The QD sample is mounted on a piezo-controlled stage under an objective lens with a numerical aperture of $NA=0.65$, giving a focal spot size of about 1\,$\upmu$m diameter. All experiments are carried out in a liquid He confocal dark-field microscope at 4.2\,K with a tunable diode laser for excitation and an avalanche photodiode (APD) for fluorescence detection. The resonant laser excitation and fluorescence detection is aligned along the same path with a microscope head that contains a 90:10 beam splitter and two polarizers. Cross-polarization enables a suppression of the spurious laser scattering into the detection path by a factor of more than $10^7$. The counts of the APD (dead time of 21.5\,ns) were binned by a QuTau time-to-digital converter with a temporal resolution of 81\,ps.

\begin{acknowledgement}
	This work was supported by the German Research Foundation (DFG) within the Collaborative Research Centre (SFB) 1242, Project No. 278162697 (TP A01), and the individual research grant No. GE2141/5-1. A.~Lu.~acknowledges gratefully support of the DFG by project LU2051/1-1 and together with A.~D.~W.~support by DFG-TRR160, BMBF - Q.Link.X 16KIS0867, and the DFH/UFA CDFA-05-06. 
\end{acknowledgement}

\begin{suppinfo}	
	Sample and device fabrication and Excitation laser intensity dependent resonance fluorescence.
\end{suppinfo}

\providecommand{\latin}[1]{#1}
\makeatletter
\providecommand{\doi}
{\begingroup\let\do\@makeother\dospecials
	\catcode`\{=1 \catcode`\}=2 \doi@aux}
\providecommand{\doi@aux}[1]{\endgroup\texttt{#1}}
\makeatother
\providecommand*\mcitethebibliography{\thebibliography}
\csname @ifundefined\endcsname{endmcitethebibliography}
{\let\endmcitethebibliography\endthebibliography}{}

\end{document}